\documentclass[journal]{IEEEtran}
\usepackage{cite}
\usepackage{amsmath,amssymb,amsfonts}
\usepackage{subfigure}
\usepackage{algorithmic}
\usepackage{algorithm}
\usepackage{graphicx}
\usepackage{textcomp}
\def\BibTeX{{\rm B\kern-.05em{\sc i\kern-.025em b}\kern-.08em
    T\kern-.1667em\lower.7ex\hbox{E}\kern-.125emX}}
\usepackage[colorlinks,linkcolor=red]{hyperref}
\usepackage{booktabs}
\usepackage{threeparttable}
\usepackage{multicol}
\usepackage{multirow}
\usepackage{xspace}
\usepackage{bm}
\usepackage{balance}
\usepackage{color}
\definecolor{qiu}{RGB}{250,10,10}

\def\ie{\emph{i.e.,~}}
\def\eg{\emph{e.g.,~}}

\def\wrt{\emph{w.r.t.~}}
\def\etal{\emph{et al.~}}

\definecolor{fan}{RGB}{181,68,52}

\def\cpar{\hss\egroup\line\bgroup\hss}

\newcommand{\ournet}{COVID-DA\xspace}

  \def\mD{{\mathcal D}}

  \def\mL{{\mathcal L}}

  \DeclareMathAlphabet\mathbfcal{OMS}{cmsy}{b}{n}
  \def\0{{\bf 0}}
  \def\1{{\bf 1}}

  \def\bff{{\bf f}}

  \def\bx{{\bf x}}
  \def\by{{\bf y}}

  \def\mmP{{\mathrm P}}

  \def\bx{{\bf x}}
  
  \def\by{{\bf y}}

  \def\citep{\cite}
  \def\etal{{\em et al.\/}\, }

\begin{document}
\title{COVID-DA: Deep Domain Adaptation from \\ Typical Pneumonia to COVID-19}
\author{Yifan Zhang, Shuaicheng Niu,  Zhen Qiu, Ying Wei,  Peilin Zhao, Jianhua Yao, \\Junzhou Huang, Qingyao Wu, and Mingkui Tan
\thanks{This work was partially supported by Key-Area Research and Development Program of Guangdong Province (2018B010107001, 2018B010108002, 2019B010155002, 2019B010155001), National Natural Science Foundation of China (NSFC) 61836003 (key project), 61876208, 2017ZT07X183, Tencent AI Lab Rhino-Bird Focused Research Program (No.JR201902), Fundamental Research Funds for the Central Universities D2191240, Pearl River S\&T Nova Program of Guangzhou 201806010081.}
\thanks{Y. Zhang, S. Niu, Z. Qiu, Q. Wu and M. Tan are with South China University of Technology, Guangzhou 510641, China, and also with Guangzhou Laboratory, Guangzhou 510335, China (e-mail: \{sezyifan, sensc, seqiuzhen\}@mail.scut.edu.cn; \{qyw, mingkuitan\}@scut.edu.cn).}
%\thanks{Q. Wu is with South China University of Technology, Guangzhou 510641, China (email: qyw@scut.edu.cn).}
\thanks{Y. Wei, P. Zhao, J. Yao and J. Huang are with Tencent AI Lab, Shenzhen 518000, China (email: \{judywei, jianhuayao, masonzhao, joehhuang\}@tencent.com).}
\thanks{Corresponding to Mingkui Tan.}
\thanks{This work is finished when Yifan Zhang and Shuaicheng Niu work as interns in Tencent AI Lab.}
}

\maketitle
\begin{abstract}
The outbreak of novel coronavirus disease 2019 (COVID-19) has already infected millions of people and is still rapidly spreading all over the globe. Most COVID-19 patients suffer from lung infection, so one important diagnostic method is to screen chest radiography images, \eg X-Ray or CT images. However, such examinations are time-consuming and labor-intensive, leading to limited diagnostic efficiency. To solve this issue, AI-based technologies, such as deep learning, have been used recently as effective computer-aided means to improve diagnostic efficiency. However, one practical and critical difficulty is the limited availability of annotated COVID-19 data, due to the prohibitive annotation costs and urgent work of doctors to fight against the pandemic. This makes the learning of deep diagnosis models very challenging. To address this, motivated by that typical pneumonia has similar characteristics with COVID-19 and many pneumonia datasets are publicly available, we propose to conduct domain knowledge adaptation from typical pneumonia to COVID-19. There are two main challenges: 1) the discrepancy of data distributions between domains; 2) the task difference between the diagnosis of typical pneumonia and COVID-19. To address them, we propose a new deep domain adaptation method for COVID-19 diagnosis, namely COVID-DA. Specifically, we alleviate the domain discrepancy via feature adversarial adaptation and handle the task difference issue via a novel classifier separation scheme. In this way, \ournet is able to diagnose COVID-19 effectively with only a small number of COVID-19 annotations. Extensive experiments verify the effectiveness of \ournet and its great potential for real-world applications.
\end{abstract}


\vspace{0.2in}
%\newpage
\section{Introduction}\label{sec:intro}
\IEEEPARstart{T}{he}
outbreak of novel coronavirus disease 2019 (COVID-19) has rapidly spread worldwide~\cite{wu2020nowcasting,wang2020novel}. To date (April 28, 2020) there have been 2,954,222 confirmed cases of COVID-19 (including 202,597 deaths, with a fatal rate of 6.9\%)~\cite{whoreport48}, leading to great threats for global public health. Due to COVID-19, many countries have been forced into emergencies~\cite{sohrabi2020world} and suffered from devastating effects on population health~\cite{pan2020clinical} and social economy~\cite{fernandes2020economic,baldwin2020thinking}.
To fight against COVID-19, one key step is to diagnose patients and  provide immediate medical treatment, thereby preventing further spread of COVID-19.

Currently, the main diagnosis method for COVID-19 is the real-time Reverse-Transcriptase Polymerase Chain Reaction (RT-PCR)~\cite{chan2020familial} test, which is regarded as the gold standard for COVID-19 detection. However, RT-PCR has a lower diagnostic sensitivity and generally requires repeated tests for the final confirmation of infection~\cite{fang2020sensitivity}. As a result, the RT-PCR test is very time-consuming and laborious~\cite{wang2020covid}. Meanwhile, it is extremely difficult for hospitals in hyper-endemic regions to provide sufficient RT-PCR tests for tons of suspected patients. To conquer this issue, an alternative diagnostic method is based on the screening of chest radiography images (CRIs), \eg X-ray or computed
tomography (CT) images~\cite{zhang2020covid}, since COVID-19 patients  often present abnormal characteristics of lung infection on CRIs~\cite{huang2020clinical,ng2020imaging}.  Compared with RT-PCR, CRI-based  diagnosis method is more efficient and has been widely used in clinical diagnosis in practice~\cite{shi2020review,shi2020large}. Nevertheless, when dealing with tons of patients, medical specialists still need to screen CRIs one by one, which, however, is highly stressful and time-consuming. Hence, there is an urgent need to develop computer-aided methods for the diagnosis of COVID-19, which help to improve the diagnostic efficiency of medical specialists.

Recently, deep learning (DL) has achieved remarkable success in medical image analysis~\cite{gomez2019deep,wang2019ct,ariz2018dynamic,anthimopoulos2016lung,tajbakhsh2016convolutional,gupta2018cnn,niu2020disturbance}.
It is a natural idea to develop DL-based methods for COVID-19 diagnosis. One of the key factors behind the success of DL is the large amount of labeled data~\cite{cao2019learning}. However, in the diagnosis of COVID-19, such extensive annotations are unavailable now due to prohibitive annotation costs and urgent work of doctors to fight against the pandemic. Hence, there is a strong motivation to develop domain adaptation~\cite{pan2009survey} to improve DL-based diagnostic models for COVID-19. Specifically, domain adaptation leverages a source domain with rich labeled data to help the model learning on the target domain. Considering that typical pneumonia has some similar characteristics with COVID-19 and many open-source pneumonia datasets are accessible, we seek to conduct domain adaptation from typical pneumonia to COVID-19 in this paper.

\clearpage
In this task, there are two major challenges. The first one is the \textbf{domain discrepancy} of data distributions, which mainly derived from different medical imaging devices or techniques. In this regard, directly applying a deep model trained on the typical pneumonia domain to the COVID-19 domain tends to perform poorly and be impractical. The second challenge lies in the diagnostic \textbf{task difference} between typical pneumonia and COVID-19. The two tasks are similar but not completely the same, which may result in poor generalization of the source-trained classifier to the target domain. However, most existing domain adaptation methods~\cite{zhang2019whole,tzeng2017adversarial,tzeng2014deep} neglect the task difference issue and adopt only  one domain-shared classifier. As a result,  they may perform poorly in COVID-19 diagnosis.

To solve the above challenges, we propose a novel deep domain adaptation method for the diagnosis of COVID-19 (namely \ournet), which relies on  domain adversarial learning and a new classifier separation scheme. To be specific, we alleviate the domain discrepancy by aligning the feature distributions of two domains via feature adversarial adaptation. In this way, \ournet is able to learn domain-invariant features for classification. Based on such features, we handle the task difference issue based on a novel classifier separation scheme, which disentangles the classifier into a domain-shared classifier and two domain-specific classifiers (for two domains). Specifically, the domain-shared classifier aims to learn task-shared classification knowledge between typical pneumonia and COVID-19, while the domain-specific classifiers seek to learn task-specific classification knowledge. To this end, we train the domain-shared classifier to learn task-shared semantic information by aligning the joint distributions of two domains over features and predictions. Meanwhile,  we maximize the diversity between the domain-shared and domain-specific classifiers to make the latter ones focus on task-specific classification information. Based on the above, \ournet is able to diagnose COVID-19 effectively with only a limited amount of labeled data, and thus is more applicable in real-world applications. 
%A brief version of this paper had been published on the MICCAI conference~\cite{zhang2019whole}. Compared with it, this journal manuscript makes several significant extensions, including (1) we extend  unsupervised domain adaptation in the MICCAI version to semi-supervised domain adaptation, and simultaneously overcome a new challenge of task difference; (2) we apply the proposed method to the diagnosis of COVID-19, which is quite urgent and important now; (3) we provide more empirical studies to evaluate the proposed method.

Our main contributions are summarized as follows:
\begin{itemize}
    \item We propose a novel deep domain adaptation method for the diagnosis of COVID-19. To the best of our knowledge, this is the first attempt to study domain adaptation from typical pneumonia to COVID-19.

    \item Based on a novel classifier separation scheme and a new domain adversarial adaptation method, the proposed method is able to overcome the task difference and domain discrepancy simultaneously.

    \item We conduct extensive experiments to evaluate the proposed method. Promising results demonstrate its effectiveness and superiority, \eg the proposed method improves the diagnostic performacne for COVID-19 from 0.6875 to 0.9298 in terms of the F1 metric.
\end{itemize}

% Experimental results show that our method improves the COVID-19 diagnostic performance (in terms of \emph{F1 score}) from 68.75\% to 92.98\% when evaluating on 945 chest X-ray images. Note that only 77 labeled COVID-19 images are used in the training process.

The rest of this paper is organized as follows. We first present related work in Section~\ref{sec:related_work}. Following that, we detail the problem definition and the proposed method in Section~\ref{sec:method}. Next, we empirically evaluate the proposed method in Section~\ref{sec:experiments}, and conclude the paper in Section~\ref{sec:conclusion}.

%\newpage
\section{Related Work}\label{sec:related_work}

\subsection{Computer-aided Diagnosis for COVID-19}
To control the transmission of COVID-19, one of the most important steps is to screen out the infected patients, and then provide proper treatments for them. Due to the relatively low time and labour costs, chest radiography imaging (CRI), \eg X-ray or computed tomography (CT) imaging, has been widely adopted to provide diagnostic evidences for radiologists. However, when  facing tons of suspected patients, it is still time-consuming for radiologists to screen medical images one by one, leading to inferior diagnostic efficiency. To address this, based on deep learning techniques, many computer-aided diagnosis methods have been developed~\cite{shi2020review}, and some of them have been deployed in hospitals~\cite{shi2020large}. For example, Xu \etal proposed a deep learning (DL) method for the early detection of COVID-19 from Influenza-A
viral pneumonia and normal cases~\cite{xu2020deep}. Chen \etal proposed a UNet++ based deep model for segmenting the infected regions of COVID-19~\cite{chen2020deep}. Nevertheless, since deep models are notoriously data-hungry, these DL-based methods require plenty of annotated data to achieve satisfactory performance. However, in the diagnosis of COVID-19, such rich supervision is unavailable in most practical scenarios due to prohibitive annotation costs and urgent work of doctors to fight the pandemic.

To solve this issue, recent studies~\cite{wang2020covid,zhang2020covid}  directly combined publicly available typical pneumonia datasets and COVID-19 dataset together to train a multi-class classification model.  However, these methods ignore  the domain discrepancy between typical pneumonia and COVID-19, thereby resulting in limited diagnostic performance for COVID-19.
Therefore, there is an urgent need to develop task-specific domain adaptation methods for COVID-19 to improve the  performance of DL-based diagnosis models.

\begin{figure*}[ht]
\vspace{-0.2in}
\centering
\includegraphics[width=17.5cm]{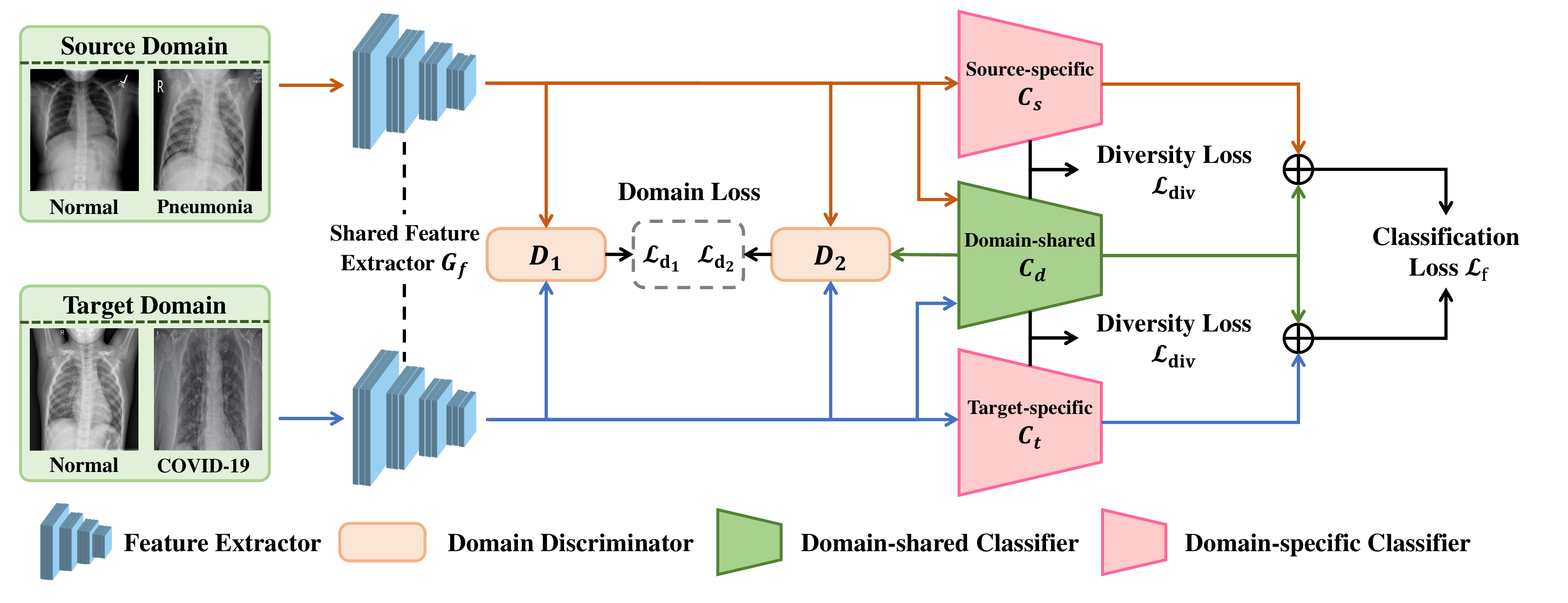}
\vspace{-0.07in}
\caption{The scheme of COVID-DA. Based on the domain discriminator $D_{1}$, we conduct feature distribution adaptation to learn domain-invariant feature extractor $G_f$ for two domains. Meanwhile, based on the discriminator $D_{2}$, we perform joint distribution adaptation for the classifier $C_{d}$ to learn domain-shared classification knowledge. We maximize the diversity between domain-shared and domain-specific classifiers to encourage the later ones to learn domain private knowledge. Moreover, we take the average ensemble (denoted by $\oplus$) of the domain-shared and domain-specific classifiers as the final prediction.}
\label{fig:overall}
\end{figure*}

\subsection{Domain Adaptation}

Most existing domain adaptation methods for natural images seek to alleviate the domain discrepancy  either by adding adaptation layers to match high-order moments of distributions, \eg DDC~\cite{tzeng2014deep}, or by devising a domain discriminator to learn domain-invariant features  in an adversarial manner, \eg DANN~\cite{ganin2014unsupervised} and MCD~\cite{saito2018maximum}. Following the latter manner, CLAN~\cite{luo2019taking}  proposed to conduct category-aware domain adaptation instead of only global alignment of domain distributions.
In medical image analysis, by taking the characteristics of medical imaging into account, Kamnitsas \etal attempted to introduce a multi-connected domain discriminator for improved adversarial training~\cite{kamnitsas2017unsupervised}. Ren \etal proposed a Siamese architecture on the target domain to add a regularization for the whole-slide images~\cite{ren2018adversarial}. However, all these methods ignore the task difference between domains, and thus may perform poorly on the  adaptation from typical pneumonia to COVID-19. To handle these, we propose a new deep domain adaptation method for COVID-19, which aims to alleviate the domain discrepancy and overcome task difference simultaneously. In this way, the proposed method is able to diagnose COVID-19 more effectively in real-world applications.

\section{Method}\label{sec:method}

\textbf{Problem Definition.}
This paper studies the problem of domain adaptation from typical Pneumonia (source domain) to COVID-19 (target domain), where the model has access to only limited labeled data from the target domain.  Formally, let $\mD^s\small{=}\{\bx_i, \by_i\}_{i=1}^{n^s}$ be  labeled source data, $\mD^t_l\small{=}\{\bx_j, \by_j\}_{j=1}^{n^t_l}$ be  limited labeled target data and $\mD^t_u\small{=}\{\bx_k\}_{k=1}^{n^t_u}$ be unlabeled target data. Here, $n^s, n^t_l, n^t_u$ denote the number of source data, labeled target data and unlabeled target data, where  $n_l^t\small{\ll} n_u^t$ and $n_l^t\small{\ll} n^s$.
Moreover, let $\mD^t\small{=}\mD^t_l\cup\mD^t_u$ be the complete target domain with  $n^t\small{=} n^t_l\small{+} n^t_u$ as the sample number.

The goal is to learn a well-performed deep model for the target domain, using both source samples (labeled) and  target samples (partially labeled).
This task, however, is very difficult due to (1) only limited labeled samples in the target domain  and (2) apparent discrepancy between typical Pneumonia and COVID-19 in terms of domain distributions and tasks. However, existing domain adaptation methods for medical images only focus on alleviating the discrepancy in terms of domain distributions, while ignoring the task difference between domains. As a result, directly applying them to the task tends to perform poorly in practice. To solve this, we propose a new deep domain adaptation method for the diagnosis of COVID-19, namely COVID-DA.

\subsection{Overall Scheme of COVID-DA}
To enforce effective domain knowledge adaptation,  we seek to alleviate the domain discrepancy  with domain adversarial adaptation and handle the task difference via a novel classifier separation scheme.
To this end, as shown in Fig.~\ref{fig:overall}, \ournet consists of three main parts: (1) a domain-shared feature extractor $G_f$ for extracting domain-invariant features; (2) two domain discriminators $\{D_{1}, D_{2}\}$ for feature adaptation and classifier adaptation, respectively; (3) a domain-shared classifier $C_{d}$  and two domain-specific classifiers $\{C_{t}, C_{s}\}$ for the diagnosis of COVID-19.  Note that the separation of domain-shared and domain-specific classifiers helps to disentangle task-shared and task-specific pathological information   regarding typical pneumonia and COVID-19.

Overall, \ournet conducts three main strategies as follows. (a) \textbf{feature adversarial adaptation:} we impose a domain loss $\mL_{\mathrm{d}_1}$ to align the feature distributions of two domains, so that the domain discrepancy is minimized  in an adversarial learning manner~\cite{tzeng2017adversarial,zhang2019whole};  (b) \textbf{classifier adversarial adaptation:} we exploit a domain loss $\mL_{\mathrm{d}_2}$ to conduct joint distribution alignment  for the domain-shared classifier $C_{d}$, making it able to  learn domain-shared  pathological  information   in an adversarial learning manner; (c) \textbf{classifier diversity maximization:} we maximize the diversity between the domain-shared and domain-specific classifiers via a diversity loss $\mL_{\mathrm{div}}$, so that the domain-specific classifiers can learn task-specific information in two domains. Note that the strategies (b) and (c) help \ournet  to  handle the task difference effectively. Lastly, we train the feature extractor and all classifiers via a focal classification loss $\mL_{\mathrm{f}}$~\cite{lin2017focal}, which makes the model class imbalance-aware and discriminative. In this way, \ournet is able to adapt the source domain knowledge to the target domain and diagnose COVID-19 effectively.

The overall training procedure of \ournet is to solve the following minimax problem~\cite{goodfellow2014generative}:
\begin{align}\label{eq:overall_training_problem}
    \min_{\bm{\Theta_1}} \max_{\bm{\Theta_2}} -\alpha&\underbrace{[\mL_{\mathrm{d}_1}(\theta_f,\theta_{d_1})+\mL_{\mathrm{d}_2}(\theta_{c_d},\theta_{d_2})]}_{\text{ domain loss}} \nonumber \\
    &~~~~~~~-
   \beta \underbrace{\mL_{\mathrm{div}}(\theta_{c_d}, \theta_{c_t}, \theta_{c_s})}_{\text{diversity loss}} + \underbrace{\mL_{\mathrm{f}}(\bm{\Theta}_1)}_{\text{focal loss}},
\end{align}
 where $\bm{\Theta_1} \small{=} \{\theta_f, \theta_{c_d}, \theta_{c_t}, \theta_{c_s}\}$ denotes the parameters of the feature extractor $G_f$ and all classifiers $\{C_{d}, C_{t}, C_{s}\}$,  while $\bm{\Theta_2}\small{=}\{\theta_{d_1}, \theta_{d_2}\}$ denotes the parameters of two discriminators $\{D_{1}, D_{2}\}$. Here, $\alpha$ and $\beta$ are trade-off parameters.
 %for $\mL_{\mathrm{d}}$ and $\mL_{\mathrm{div}}$, while we set the weight of $\mL_{\mathrm{f}}$ for simplicity.

We next detail two domain losses $\mL_{\mathrm{d}_1}$ and $\mL_{\mathrm{d}_2}$ in Section~\ref{sec:discriminator1} and Section~\ref{sec:discriminator2}, respectively. Following that, we detail the classifier diversity loss $\mL_{\mathrm{div}}$ in Section~\ref{sec:diversity} and then the focal loss $\mL_{\mathrm{f}}$ in Section~\ref{sec:classification}.
%where the total domain loss is the sum of two domain losses, \ie $\mL_{\mathrm{d}}= \mL_{\mathrm{d}_1}+\mL_{\mathrm{d}_2}$.

\subsection{Feature Adversarial Adaptation}\label{sec:discriminator1}%   for Domain-invariant Feature Extractor

Diverse imaging devices and preprocessing techniques intrinsically result in huge domain discrepancy in terms of data distribution. To resolve the discrepancy, we resort to domain adversarial learning for aligning feature distributions of domains. Specifically, on the one hand, a domain  discriminator $D_{1}$ is trained to adequately distinguish feature representations between two domains by minimizing a domain loss $\mL_{\mathrm{d_1}}$. On the other hand, the feature extractor $G_f$ is trained to confuse the  discriminator by  maximizing the domain loss $\mL_{\mathrm{d_1}}$. As  a  result,  the  learned  feature  extractor is able to extract domain-invariant features that confuse the discriminator well. Based on the least square distance~\cite{mao2017least}, we define the domain  loss for feature adversarial adaptation as:
\begin{align}\label{eq:domain_loss_d1}
    \mL_{\mathrm{d_1}}(\theta_f,\theta_{d_1}) = \frac{1}{n^s}\sum_{\mD^s}d_1(\bx)^2 + \frac{1}{n^t}\sum_{\mD^t}(1-d_1(\bx))^2,
\end{align}
where $d_1(\bx)=D_{1}(G_f(\bx))$ denotes the prediction of the domain discriminator \wrt $\bx$.  We set the label of the target domain to 1 and that of the source domain to 0.

\subsection{Classifier Adversarial Adaptation}\label{sec:discriminator2}

Most existing adversarial domain adaptation methods~\cite{lafarge2017domain,ganin2014unsupervised,zhang2019whole}
assume that the source domain deals with the same classification task as the target domain. Hence, they usually focus on feature distribution alignment as Section~\ref{sec:discriminator1} and use a classifier trained on the source domain to classify the target data. However, they may fail to handle the problem in this paper, since  pneumonia diagnosis and COVID-19 diagnosis are similar but not completely the same, originated from different pathological mechanisms.
Specifically, let $\mmP(\bff^s)$ and $\mmP(\bff^t)$ denote the feature distributions of the source and target domains, while let $\mmP(\by^s | \bff^s)$ and $\mmP(\by^t | \bff^t)$ be the prediction conditional distributions of two domains.  Even though the feature distributions have been matched (\ie $\mmP(\bff^s) \small{=} \mmP(\bff^t)$), the task difference potentially results in  different prediction conditional distributions (\ie $\mmP(\by^s | \bff^s) \small{\neq} \mmP(\by^t | \bff^t)$)~\cite{motiian2017unified,luo2017label}. As a result, only using the source-trained  classifier $\mmP(\by^s | \bff^s)$ may not be able to diagnose COVID-19 well.

To solve this issue, as shown in Fig.~\ref{fig:overall}, we propose a novel classifier separation scheme by disentangling the domain-shared classifier $C_{d}$ and the domain-specific classifiers $\{C_{s},C_{t}\}$.
To be specific, the domain-shared classifier seeks to acquire task-shared  classification knowledge, while the domain-specific classifiers aim to learn task-specific knowledge.
By taking the average ensemble of two classifiers as the final prediction~\cite{zhang2019collaborative}, \ournet is able to handle the task difference and diagnose COVID-19 well.

In this scheme, one key issue  is how to learn the domain-shared classifier. In fact, when the feature distributions match well  (\ie $\mmP(\bff^s) \small{=} \mmP(\bff^t)$), if we train the classifier to align the joint distributions  (\ie $\mmP(\by^s, \bff^s) \small{=} \mmP(\by^t, \bff^t)$), then the domain-shared classifier is able to learn task-shared classification knowledge since $\mmP(\by, \bff)\small{=}\mmP(\bff)\mmP(\by|\bff)$~\cite{long2017deep}. Motivated by this, we propose to train the domain-shared classifier by aligning the joint distributions via domain adversarial learning~\cite{tzeng2017adversarial}.

On the one hand, a discriminator $D_2$ is trained to adequately differentiate the joint distributions between domains by minimizing a domain loss $\mL_{\mathrm{d_2}}$. Specifically, the input of the discriminator $D_2$ consists of  both features and  predictions.   On the other hand, the domain-shared classifier $C_d$ is trained to confuse the discriminator by maximizing the domain loss. As Section~\ref{sec:discriminator1}, we define the domain loss for classifier adversarial adaptation based on the least square distance:
\begin{align}\label{eq:domain_loss_d2}
    \mL_{\mathrm{d_2}}(\theta_{c_d},\theta_{d_2}) = \frac{1}{n^s}\sum_{\mD^s}d_2(\bx)^2 + \frac{1}{n^t}\sum_{\mD^t}(1\small{-}d_2(\bx))^2,
\end{align}
where $d_2(\bx)=D_{2}\big(G_f(\bx), C_{d}(G_f(\bx))\big)$ denotes the prediction of the domain discriminator regarding the joint distribution over the feature and the prediction \wrt $\bx$.
Moreover, we denote the label of the target domain as 1 and that of the source as 0.

\subsection{Classifier Diversity Maximization}\label{sec:diversity}
In \ournet, we handle the issue of task difference by disentangling the domain-shared and domain-specific classifiers.  In Section~\ref{sec:discriminator2}, we have enforced the domain-shared classifier to acquire domain-invariant  classification knowledge. In this section, we further enforce the domain-specific classifiers to acquire domain private classification knowledge. To this end, we  maximize the diversity between the domain-specific classifier and the domain-shared classifier. Specifically, given any sample $\bx$, let
$\hat{\by}_d$ denote the prediction of domain-shared classifiers (\ie $C_{d}$), and let $\hat{\by}_t$ and $\hat{\by}_s$ denote the prediction of the domain-specific classifier (\ie $C_{t}$ and $C_{s}$). Based on the cosine distance, we maximize the classifier diversity based on the following diversity loss:
\begin{align}\label{eq:classifier_discrepancy}
    \mL_{\mathrm{div}}(\theta_{c_d}, \theta_{c_t}, \theta_{c_s}) \small{=  -}\frac{1}{n^s}\sum_{\mD^s}\mathrm{cos}\big(\hat{\by}_d,\hat{\by}_s\big)\small{-}\frac{1}{n^t}\sum_{\mD^t}\mathrm{cos}\big(\hat{\by}_d,\hat{\by}_t\big).
\end{align}

In this way, the two domain-specific classifiers are different with the domain-shared classifier as large as possible, and thus able to learn domain-specific classification information.
Moreover, since the final prediction of \ournet is the average ensemble of two classifiers, maximizing classifier diversity also enhance the performance of ensemble learning~\cite{zhou2012ensemble}.

\subsection{Focal Loss for COVID-19 Diagnosis}\label{sec:classification}

For the diagnosis of COVID-19, one can adopt any classification losses to train our \ournet, \eg cross-entropy. Nevertheless, considering the class imbalance issue in medical diagnosis, we use the focal loss~\cite{lin2017focal} as follows:
\begin{align}\label{eq:focal_target}
    \mL_{\mathrm{f}}(\bm{\Theta}_1)
    =-\frac{1}{n_l^t\small{+}n^s}\sum_{\mD_l^t\cup\mD^s}\by^\top\big((\textbf{1}\small{-}\hat{\by})^\gamma\small{\odot}\mathrm{log}(\hat{\by})\big),
\end{align}
where $\hat{\by}$ is the final prediction of \ournet \wrt a given sample $\bx$. Here, $\hat{\by}$ is the average prediction of both domain-shared ($C_d$) and domain-specific ($C_t$ or $C_s$) classifiers. Note that, the
focal loss is a widely-used loss for class imbalance issue~\cite{zhang2019whole}.
Moreover, $\odot$ denotes the element-wise product and $\gamma$ is a hyper-parameter in focal loss.

\clearpage
We summarize the training and inference details of \ournet in Algorithms~\ref{al:training} and~\ref{al:inference}, respectively. Moreover, we implement the adversarial learning via a gradient reversal layer (GRL)~\cite{lafarge2017domain,ganin2014unsupervised}, which reverses the gradient of the domain loss  when backpropagating to the feature extractor or domain-shared classifier. In this way, we are able to train \ournet through standard backpropagation in an end-to-end manner.

\begin{algorithm}[h]
    \small
    \caption{Training of \ournet}\label{al:training}
    \begin{algorithmic}[1]
    \REQUIRE Labeled source data $\mD^s\small{=}\{\bx_i, \by_i\}_{i=1}^{n^s}$, labeled target data $\mD^t_l\small{=}\{\bx_j, \by_j\}_{j=1}^{n^t_l}$ and unlabeled target data $\mD^t_u\small{=}\{\bx_k\}_{k=1}^{n^t_u}$; Training epoch $M$; Trade-off hyper-parameters $\alpha$ and $\beta$.
    \ENSURE Parameters of \ournet: Feature extractor $G_f$; classifiers $\{C_{d}, C_{t}, C_{s}\}$; domain discriminators $\{D_{1}, D_{2}\}$.
    \FOR{$m = 1 \to M$}
        \STATE Extract feature vector $\bff$ based on $G_f$;
        \STATE Obtain the predictions $\{\hat{\by}_d,\hat{\by}_t,\hat{\by}_s\}$ of  $\{C_{d},C_{t},C_{s}\}$ based on $\bff$, respectively;
        \STATE Compute the domain adversarial losses $\mL_{\mathrm{d}_1}$ and $\mL_{\mathrm{d}_2}$ based on $\bff$ and $\{C_{d},D_{1},D_{2}\}$; // Eqn.~(\ref{eq:domain_loss_d1}) and Eqn.~(\ref{eq:domain_loss_d2})
        \STATE Compute the classifier diversity loss $\mL_{\mathrm{di v}}$ based on $\{\hat{\by}_d,\hat{\by}_t,\hat{\by}_s\}$; // Eqn.~(\ref{eq:classifier_discrepancy})
        \IF{labeled data}
        \STATE Compute the classification loss  $\mL_{\mathrm{f}}$ based on $\{\hat{\by}_d,\hat{\by}_t\}$ and $\{\hat{\by}_d,\hat{\by}_s\}$.~~// Eqn.~(\ref{eq:focal_target})
        \ENDIF
        \STATE Compute the total loss;~~// Eqn.~(\ref{eq:overall_training_problem})
        \STATE loss.backward(). ~~~//  standard backward propagation
    \ENDFOR
    \RETURN $G_{f}$, $C_{d}$ and $C_{t}$.
     \end{algorithmic}
\end{algorithm}

\begin{algorithm}
    \small
    \caption{Inference of \ournet}\label{al:inference}
    \begin{algorithmic}[1]
    \REQUIRE Target COVID-19 data $\bx$; Parameters of feature extractor $G_{f}$; Two  classifiers for the target domain $\{C_{d},C_{t}\}$.
    \STATE Extract feature $\bff$ regrading $\bx$ using $G_f$;
    \STATE Compute the predictions \{$\hat{\by}_d, \hat{\by}_t\}$ using $\{C_{d}, C_{t}\}$ based on $\bff$;
    \STATE Compute the ensemble prediction $\hat{\by}=\big(\hat{\by}_d$+$\hat{\by}_t\big)/2$.
    \RETURN $\hat{\by}$.
    \end{algorithmic}
\end{algorithm}

\section{Experimental Results}\label{sec:experiments}

To verify the proposed method\footnote{We will make the source code publicly available.}, we evaluate \ournet on two main aspects: (1) the performance in the diagnosis of COVID-19; (2) the algorithm characteristics of \ournet.

\subsection{Experimental Settings}
\subsubsection{Dataset}
The dataset\footnote{The dataset is available at https://github.com/qiuzhen8484/COVID-DA.git.} used in this experiment is collected from three open-source datasets,
\ie the COVID chest X-ray dataset~\cite{cohen2020covid}, the COVID-19 Radiography Database\footnote{https://www.kaggle.com/tawsifurrahman/covid19-radiography-database.} and the dataset of RSNA Pneumonia Detection Challenge on Kaggle\footnote{https://www.kaggle.com/c/rsna-pneumonia-detection-challenge/data.}. Based on these collected data, we randomly choose part of normal cases and all typical pneumonia cases to make up the source domain, and use the rest of normal cases and all COVID-19 cases as the target domain.
The statistics of two domains are summarized in Table~\ref{tab:data}.

Note that \textbf{only 30\% of the training COVID-19 samples are labeled in the training process}, which would be more practical in real-world scenarios.
Moreover, the three datasets are acquired from different countries with various imaging devices, while the tasks of two domains are similar but not completely the same. Therefore, this domain adaptation task suffers from severe domain discrepancy and  task difference. In addition, as shown in Table~\ref{tab:data}, the class imbalance is also severe. Considering the above issues, such a diagnosis task of COVID-19 is very challenging.

\begin{table}[t]
  \caption{Statistics of the dataset, where pneumonia serves as the source domain and COVID-19 serves as the target domain.}
  \vspace{-0.03in}
  \centering
  \begin{threeparttable}
  \renewcommand\arraystretch{0.7}
  \renewcommand{\tabcolsep}{2.5pt}
  \resizebox{0.46\textwidth}{!}{
  \label{tab:data}
    \begin{tabular}[width=\textwidth]{cccccccccc}
    \toprule
    \multirow{2}{*}{Set} &\multirow{2}{*}{Domain}&
    \multicolumn{3}{c}{Categories}&
    \multirow{2}{*}{\#Total}\cr
    \cmidrule(lr){3-5}
    & &  \#Normal & \#Pneumonia & \#COVID-19& & \cr
    \midrule
    \multirow{2}{*}{Training} & Pneumonia  & 5613 & 2306 & 0 & 7919\cr
     & COVID-19 & 2541 & 0 & 258 & 2799\cr
    \midrule
    Test& COVID-19 & 885 & 0 & 60 & 945\cr
    \bottomrule
    \end{tabular}
  }
  \vspace{-0.1in}
    \end{threeparttable}
\end{table}

\subsubsection{Compared methods}
We compare \ournet with four categories of methods. (1) \textbf{Baselines}:   Source-only (training the model on the well-labeled source domain), Target-only (training the model on the target domain with limited labels) and Fine-tuning (training the model on the well-labeled source domain and then fine-tuning it on the target domain with limited labels); (2) \textbf{deep diagnostic models} for COVID-19:  DLAD~\cite{zhang2020covid} and COVID-Net~\cite{wang2020covid} (training on the labeled target domain); (3) \textbf{Unsupervised domain adaptation}: MCD~\cite{saito2018maximum}, DSN~\cite{bousmalis2016domain}, DANN~\cite{ganin2014unsupervised}, DMAN~\cite{zhang2019whole}, which train deep models on both the labeled source domain and unlabeled target domain; (4) \textbf{Semi-supervised domain adaptation}: SDT~\cite{7410820} and semi-DMAN (extended from DMAN~\cite{zhang2019whole}). Note that, the semi-supervised domain adaptation methods train deep models using both labeled source data and partially-labeled target data.

\begin{table*}[ht]
\centering
\caption{Comparisons on COVID-19 diagnosis in terms of F1 score (\%), Recall (\%), Precision (\%), AUC, Sum (\%) and Cost.}
\renewcommand\arraystretch{0.9}
\resizebox{0.8\textwidth}{!}{
% \scalebox{1.2}{
\label{tab:main-result}
\begin{tabular}{ccccccc}
\toprule
~~Method~~ & ~~F1 ($\uparrow$)~~  & ~Recall ($\uparrow$)~ & Precision ($\uparrow$) & ~~AUC ($\uparrow$)~~ & ~~Sum ($\uparrow$)~~ & ~~Cost ($\downarrow$)~~  \\
\midrule
Source-only  & 65.04 & 66.67 & 63.49 & 0.899 & 82.03 & 20.3  \\
Target-only  & 68.75 & 55.00 & 91.67 & 0.971 & 77.33 & 24.6  \\
Fine-tuning  & 64.29 & 75.00 & 56.25 & 0.946 & 85.52 & 17.0  \\
DLAD~\cite{zhang2020covid}  & 67.18 & 73.33 & 61.97 & 0.961 & 85.14 & 17.1  \\
COVID-Net~\cite{wang2020covid}  & 71.94 & 83.33 & 63.29 & 0.977 & 90.03 & 11.9  \\
MCD~\cite{saito2018maximum}  & 61.54 & 60.00 & 63.16 & 0.904 & 78.81 & 23.7  \\
DANN~\cite{ganin2014unsupervised}  & 66.15 & 71.67 & 61.43 & 0.904 & 84.31 & 18.0  \\
DSN~\cite{bousmalis2016domain}  & 73.02 & 76.67 & 69.70 & 0.884 & 87.20 & 14.6  \\
DMAN~\cite{zhang2019whole}  & 75.63 & 75.00 & 76.27 & 0.915 & 86.71 & 14.9  \\
Semi-DMAN  & 77.27 & 85.00 & 70.83 & 0.978 & 91.31 & 10.2  \\
SDT~\cite{7410820}  & 79.69 & 85.00 & 75.00 & 0.962 & 91.54 & 9.8  \\
\midrule
COVID-DA  & {\bf 92.98} & {\bf 88.33} & {\bf 98.15} & {\bf 0.985} & {\bf 94.11} & {\bf 6.4}  \\
\bottomrule
\end{tabular}
}
\end{table*}

% \begin{table*}[h]
% \centering
% \caption{Ablation studies on COVID-DA in terms of six metrics. Both cross entropy loss $\mL_{\mathrm{ce}}$ and focal loss $\mL_{\mathrm{f}}$ are \\ used for classification; Domain losses $\mL_{\mathrm{d_{1}}}$ and  $\mL_{\mathrm{d_{2}}}$ are used for domain adaptation regarding discriminators $D_{1}$ and $D_2$;  \\ Diversity loss $\mL_{\mathrm{div}}$ is employed  to maximize the classifier diversity.}
% % \renewcommand\arraystretch{0.8}
%   \renewcommand{\tabcolsep}{4.0pt}
% % \scalebox{1.1}{
% \resizebox{0.8\textwidth}{!}{
% \label{module-eval}
% \begin{tabular}{cccccc|cccccc}
% \toprule
% Backbone &$\mL_{\mathrm{ce}}$ & $\mL_{\mathrm{f}}$ & $\mL_{\mathrm{d_{1}}}$ & $\mL_{\mathrm{d_{2}}}$ & $\mL_{\mathrm{div}}$ & F1 ($\uparrow$)  & Recall ($\uparrow$) & Precision ($\uparrow$) & AUC ($\uparrow$) & Sum ($\uparrow$)& Cost ($\downarrow$)\\
% \midrule
% $\surd$ &$\surd$& &&& & 85.98 & 76.67 & 97.87 & 0.991 & 88.28 & 12.7  \\
% $\surd$& &$\surd$&&& & 87.04 & 78.33 & 97.92 & 0.992 & 89.11 & 11.8  \\
% % $\surd$ &&$\surd$&&& & 87.85 & 78.33 & {\bf 100.00} & 89.17 & 11.7 & 0.9945 \\
% $\surd$ &&$\surd$&$\surd$&& & 88.89 & 80.00 & {\bf 100.00} & 0.988 & 90.00 & 10.8  \\
% $\surd$ &&$\surd$&$\surd$&$\surd$& & 91.07 & 85.00 & 98.08 & {\bf 0.996} & 92.44 & 8.2  \\
% \midrule
% $\surd$ &&$\surd$&$\surd$&$\surd$& $\surd$& {\bf 92.98} & {\bf 88.33} & 98.15 & 0.985 & {\bf 94.11} & {\bf 6.4} \\
% \bottomrule
% \end{tabular}
% }
% \end{table*}

\begin{table*}[ht]
\centering
\caption{Ablation studies on COVID-DA in terms of six metrics. Both cross entropy loss $\mL_{\mathrm{ce}}$ and focal loss $\mL_{\mathrm{f}}$ are used for classification; Domain losses $\mL_{\mathrm{d_{1}}}$ and  $\mL_{\mathrm{d_{2}}}$ are used for domain adaptation regarding discriminators $D_{1}$ and $D_2$;  Diversity loss $\mL_{\mathrm{div}}$ is employed  to maximize the classifier diversity.}
\renewcommand\arraystretch{0.9}
  \renewcommand{\tabcolsep}{4.0pt}
% \scalebox{1.1}{
\resizebox{0.8\textwidth}{!}{
\label{module-eval}
\begin{tabular}{cccccc|cccccc}
\toprule
Backbone &$\mL_{\mathrm{ce}}$ & $\mL_{\mathrm{f}}$ & $\mL_{\mathrm{d_{1}}}$ & $\mL_{\mathrm{d_{2}}}$ & $\mL_{\mathrm{div}}$ & F1 ($\uparrow$)  & Recall ($\uparrow$) & Precision ($\uparrow$) & AUC ($\uparrow$) & Sum ($\uparrow$)& Cost ($\downarrow$)\\
\midrule
$\surd$ &$\surd$& &&& & 70.71 & 58.33 & 89.74 & 0.986 & 78.94 & 22.9  \\
$\surd$& &$\surd$&&& & 72.92 & 58.33 & 97.22 & 0.985 & 79.11 & 22.6  \\
$\surd$ &&$\surd$&$\surd$&& & 84.62 & 73.33 & {\bf 100.00} & 0.974 & 86.67 & 14.4  \\
$\surd$ &&$\surd$&$\surd$&$\surd$& & 91.07 & 85.00 & 98.08 & {\bf 0.996} & 92.44 & 8.2  \\
\midrule
$\surd$ &&$\surd$&$\surd$&$\surd$& $\surd$& {\bf 92.98} & {\bf 88.33} & 98.15 & 0.985 & {\bf 94.11} & {\bf 6.4} \\
\bottomrule
\end{tabular}
}
\end{table*}

\subsubsection{Implementation details}\label{sec:inplementation_details}
We implement our method based on PyTorch~\cite{paszke2019pytorch}. For a fair comparison, we adopt a Resnet-18~\cite{he2016deep} model, pretrained on ImageNet~\cite{deng2009imagenet}, as the backbone of all methods. For all compared methods, we keep the same hyper-parameters as the original paper. For COVID-DA, the details consist of two parts. (1) \textbf{network architectures}: We implement the feature extractor based on Resnet-18, while we implement the two domain discriminators based on two-layer fully-connected networks~\cite{zhang2019whole}, and implement all classifiers by one fully-connected layer; (2) \textbf{parameter settings}: In the training process, we use an SGD optimizer with the learning rate of 0.001 to train the whole network. The batch size for each domain is set to 16. As for the trade-off parameters in Eqn.~(\ref{eq:overall_training_problem}), we set $\alpha\small{=}0.1$ and $\beta\small{=}0.1$ via cross-validation. Following~\cite{lin2017focal}, we set $\gamma\small{=}2.0$ for the focal loss.

\subsubsection{Evaluation metrics}
We evaluate the  diagnostic performance of all considered methods for COVID-19 in terms of six metrics: \emph{F1 score} (\%), \emph{Recall} (\%), \emph{Precision} (\%), \emph{AUC}, \emph{Sum} (\%) and  \emph{Cost}. Specifically, the first three metrics are calculated based on previous studies~\cite{zhang2019whole,zhang2019collaborative}, the \emph{AUC} metric is based on the work~\cite{ding2017large}, and the \emph{Sum} and \emph{Cost} metrics are based on~\cite{zhang2018online,zhang2019online}.  We recommend readers to these papers for detailed implementations of these metrics.

\subsection{Evaluation on COVID-19 Diagnosis}
\subsubsection{Results}
We evaluate all methods in terms of six metrics and report the results in Table~\ref{tab:main-result}. \textbf{Overall},  the proposed method performs the best, which confirms its effectiveness and superiority in COVID-19 diagnosis. Since there is an urgent need for computer-aided diagnosis for COVID-19, \ournet makes great medical sense in practice.

According to Table~\ref{tab:main-result}, we draw the following observations. \textbf{(1)} Both Target-only and Fine-tuning do not certainly outperform the Source-only on all considered evaluation metrics. One possible reason is that the deep model may overfit to the limited labeled data of the target domain, and thus perform limitedly on the test set. \textbf{(2)} deep diagnostic models (DLAD and COVID-Net) outperform Target-only, which demonstrates the contribution of particularly devised architectures. \textbf{(3)} most unsupervised domain adaptation (DA) methods (\eg DANN, DSN and DMAN)  outperform Source-only, which confirms the effectiveness of DA.
\textbf{(4)} The semi-supervised DA methods (\ie SDT and semi-DMAN) further improve the model performance on the target domain. This result verifies the contribution of limited target annotations in DA. \textbf{(5)} \ournet outperforms all compared methods, which demonstrates the superiority of the proposed method to leverage both well-labeled source data and partially-labeled target data.

\subsubsection{Visualization of Grad-CAMs}

\begin{figure}[h]
\centering
\includegraphics[width=0.48\textwidth]{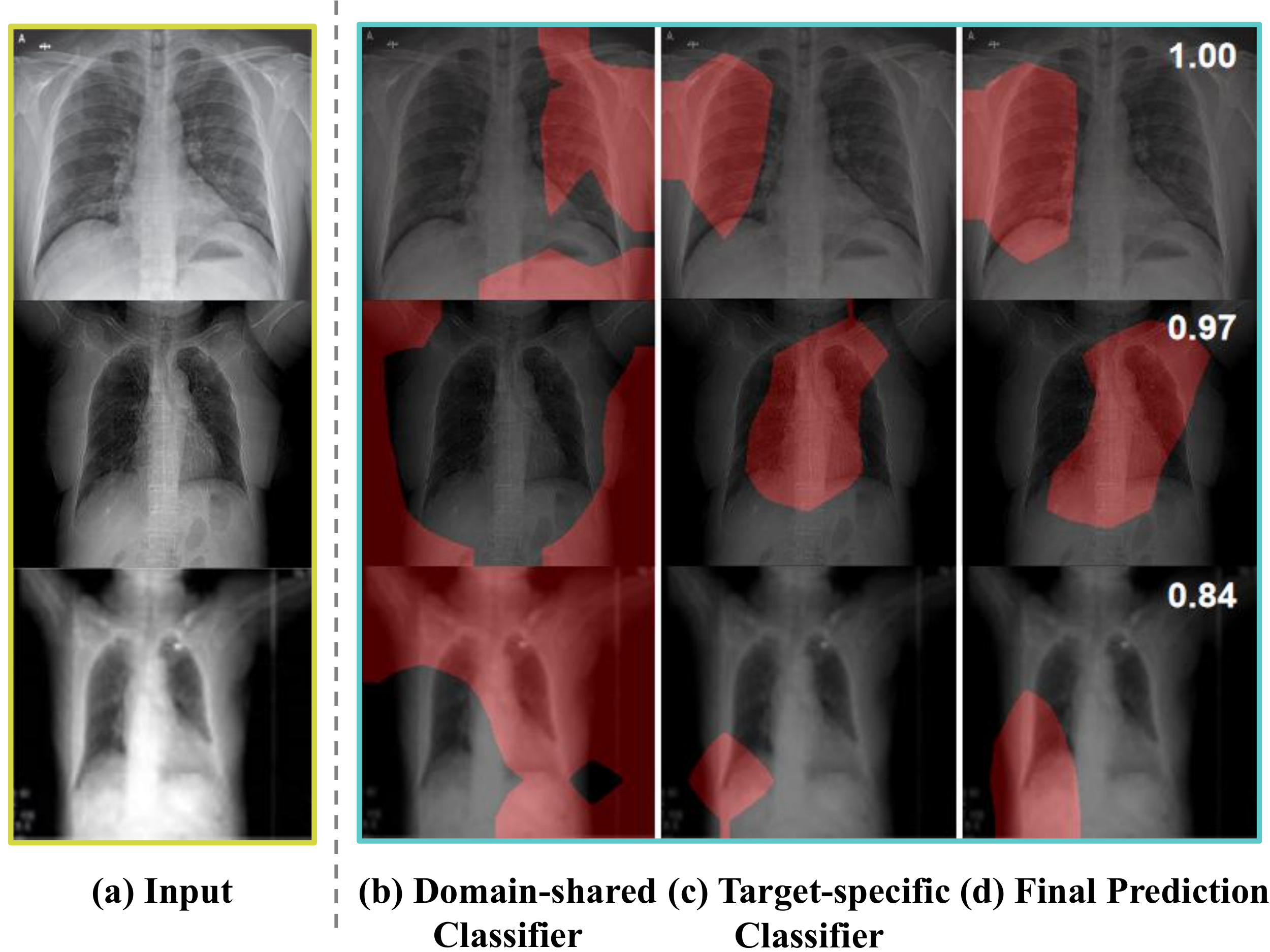}
\caption{Visualization of three COVID-19 confirmed patients' chest X-ray images and the corresponding Grad-CAMs obtained by our method. (a) the original input image; (b) the Grad-CAMs obtained by our domain-shared classifier; (c) the Grad-CAMs obtained by our target-specific classifier; (d) the Grad-CAMs obtained by the ensemble of both domain-shared and target-specific classifiers. Specifically, the red-coloured area is the region where the classifiers focus on. The probabilities on the rightest column denote the confidence of the final prediction.}
\label{fig:COVID}
\end{figure}

\begin{table*}[ht]
  \centering
%   \begin{threeparttable}
  \caption{Influence of the trade-off parameters $\alpha$ and $\beta$ in \ournet. Here, $\alpha$ and $\beta$ control the domain losses ($\mL_{\mathrm{d}_1}$ and $\mL_{\mathrm{d}_2}$) and classifier diversity loss ($\mL_{\mathrm{div}}$), respectively.}
  \label{tab:parameter_sensiti}
%   \renewcommand\arraystretch{0.9}
%   \renewcommand{\tabcolsep}{4.0pt}
%   \scalebox{1.0}{
    \resizebox{0.8\textwidth}{!}{
    \begin{tabular}[width=\textwidth]{cccccccc}
    \toprule
   Parameter &Value & ~~F1 ($\uparrow$)~~  & ~Recall ($\uparrow$)~ & Precision ($\uparrow$) & ~~AUC ($\uparrow$)~~ & ~~Sum ($\uparrow$)~~ & ~~Cost ($\downarrow$)~~ \cr
    \midrule
    \multirow{4}{*}{$\alpha$}&
    $10^{-3}$ &90.09&83.33&98.04 & 0.961 & 91.61 & 9.1 \cr
      &$10^{-2}$ &91.89&85.00&{\bf100.00} &{\bf0.990} &92.50 & 8.1 \cr
     &$10^{-1}$&{\bf92.98}&{\bf88.33}&98.15 & 0.985 & {\bf94.11} & {\bf6.4} \cr
     &$1$ & 87.27 & 80.00 & 96.00 & 0.980 & 89.89 & 11\cr
     \midrule
   \multirow{4}{*}{$\beta$}&
    $10^{-3}$ & 89.47 & 85.00 & 94.44 & 0.984 & 92.33 & 8.4\cr
    &$10^{-2}$ & 90.09 & 83.33 & 98.04 & {\bf0.985} & 91.61 & 9.1 \cr
     &$10^{-1}$ & {\bf92.98} & {\bf88.33} & {\bf98.15} &{\bf0.985} & {\bf94.11} & {\bf6.4} \cr
     &$1$ & 90.43 & 86.67 & 94.55 & 0.976 & 93.16 & 7.5 \cr
    \bottomrule
    \end{tabular}
    }
    % }
    % \end{threeparttable}
\end{table*}

\begin{table*}[ht]
  \centering
%   \begin{threeparttable}
  \caption{
  Comparisons of different measurements for computing domain adversarial losses ($\mL_{\mathrm{d}_1}$ and $\mL_{\mathrm{d}_2}$)  and the classifier diversity loss ($\mL_{\mathrm{div}}$) in \ournet.}
  \label{tab:loss}
    \newcommand{\tabincell}[2]{\begin{tabular}{@{}#1@{}}#2\end{tabular}}
%   \renewcommand\arraystretch{0.95}
%   \renewcommand{\tabcolsep}{5.0pt}
%   \scalebox{1.0}{
    \resizebox{0.8\textwidth}{!}{
    \begin{tabular}{cccccccc}
    \toprule
   Objective &Measurement & F1 ($\uparrow$)  & Recall ($\uparrow$) & Precision ($\uparrow$) & AUC ($\uparrow$) & Sum ($\uparrow$)& Cost ($\downarrow$)\cr
    \midrule
       \multirow{3}{*}{\tabincell{c}{Domain Losses \\ $(\mL_{\mathrm{d}_1}$~and~$\mL_{\mathrm{d}_2})$}}
    & Focal Loss &87.85&78.33&{\bf 100.00} &0.984 &89.17  & 11.7  \cr
    &GAN Loss & 89.91 & 81.67 & {\bf 100.00} & {\bf0.988} & 90.83  & 9.9  \cr
     &Least Square Loss & {\bf 92.98} & {\bf 88.33} & 98.15 & 0.985 & {\bf 94.11}  & {\bf 6.4}  \cr
     \midrule
     \multirow{5}{*}{\tabincell{c}{Diversity Loss\\ $(\mL_{\mathrm{div}})$}}&
    L1 Distance & 91.38 & {\bf88.33} & 94.64 & 0.963 & 94.00 & 6.6  \cr
      &L2  Distance & 90.09 & 83.33 & 98.04 & 0.963 & 91.61  & 9.1  \cr
     &KL Divergence & 85.71 & 85.00 & 86.44 & 0.950 & 92.05  & 8.9  \cr
     &JS Divergence & 90.27 & 85.00 & 96.23 & 0.977 & 92.39 & 8.3  \cr
      &Cosine Distance & {\bf92.98} & {\bf88.33} & {\bf98.15} & {\bf 0.985} & {\bf94.11} & {\bf6.4} \cr
    \bottomrule
    \end{tabular}
    }
    % }
    % \end{threeparttable}
\end{table*}

In this section, we use the Gradient-weighted Class Activation Mapping~\cite{selvaraju2017grad} (Grad-CAM) method  to visualize the important regions that the devised classifiers focus on for predictions. We present the visualization results of three COVID-19 patients in Fig.~\ref{fig:COVID}. As expected, the domain-shared classifier focuses more on the surrounding regions to capture the task-shared classification information, while the target-specific classifier focuses more on pathological regions of COVID-19 itself. By combining the above two classifiers, \ournet is able to make predictions based on both task-shared and target-specific classification information. In addition, the visualized Grad-CAMs are also an interpretation for the prediction of \ournet, which helps doctors to judge the prediction reliability in practice.

% \newpage
\subsubsection{Ablation studies}
We conduct ablation studies to evaluate the effectiveness of different components in \ournet. As shown in Table~\ref{module-eval}, all components in our methods (\ie feature adversarial adaptation, classifier adversarial adaptation, classifier diversity maximization, and focal loss) make empirical contributions and play important roles in our method. Particularly, domain adversarial losses ($\mL_{\mathrm{d_1}}$ and $\mL_{\mathrm{d_2}}$) are relatively important. Moreover, the classifier diversity loss ($\mL_{\mathrm{div}}$) is able to further improve  diagnostic performance. These results demonstrate the necessity to reduce the domain discrepancy and overcome the task difference in the task of domain adaptation for COVID-19.

\subsection{More Discussions}

\subsubsection{Parameter sensitivities}\label{sec:parameter_analysis}
As mentioned in Section~\ref{sec:inplementation_details}, we set trade-off parameters $\alpha=0.1$ and $\beta=0.1$ in all experiments, where $\alpha$ adjusts the domain adversarial losses ($\mL_{\mathrm{d}_1}$ and $\mL_{\mathrm{d}_2}$) and $\beta$ controls the classifier diversity loss ($\mL_\mathrm{div}$). In this section, we evaluate the sensitivities of these two parameters, where we only evaluate one parameter each time, fixing all other parameters. From Table~\ref{tab:parameter_sensiti}, our proposed method achieves the best or relatively good performance with the setting $\alpha=0.1$ and $\beta=0.1$.

\subsubsection{Domain adversarial loss}

In our method, we define the domain adversarial losses (in Eqns.~(\ref{eq:domain_loss_d1}) and~(\ref{eq:domain_loss_d2})) relying on the least square loss, since it helps to improve domain confusion and
stabilize training, by preserving the domain distance information~\cite{mao2017least}. In this section, we empirically compare it with the focal loss and original GAN loss~\cite{goodfellow2014generative}. As shown in Table~\ref{tab:loss}, the least-square loss outperforms another two losses, which demonstrates the superiority of the adopted domain loss.

\subsubsection{Classifier diversity loss}
In our method, we define the classifier diversity loss (in Eqn.~(\ref{eq:classifier_discrepancy})) relying on the cosine distance. In fact, one can also use other distance metrics, \eg  L1 distance, L2 distance, KL divergence and JS divergence, according to the tasks at hand. To find the most suitable one, we conduct many preliminary experiments and report the results in Table~\ref{tab:loss}.
To be specific, the cosine distance performs the best over all considered distances in the domain adaptation task for COVID-19.

\section{Conclusion}\label{sec:conclusion}
In this paper, we have proposed a deep domain adaptation method for the diagnosis of COVID-19 (namely \ournet), which aims to transfer the domain knowledge from the well-labeled source domain (\ie typical pneumonia) to the partially-labeled target domain (\ie COVID-19). To be specific, we minimize the domain discrepancy by  aligning the feature distributions of two domains via domain adversarial learning. Meanwhile, we develop a novel classifier separation scheme to overcome the issue of task difference between domains. In this way, the proposed method is able to learn a well-performed deep model with very limited annotations of COVID-19. Extensive experiments demonstrate the effectiveness and superiority of \ournet.

It is worth mentioning that our proposed method is of great clinical importance, since extensive annotations of COVID-19 are inaccessible now, while there is an urgent demand to develop deep learning based diagnosis methods for COVID-19. In the future, one can apply \ournet to CT imaging-based diagnosis of COVID-19 and extend it to the segmentation task for fine-grained diagnosis of COVID-19.

\newpage
\bibliographystyle{IEEEtran}
{
	\balance
	\bibliography{reftmi}
}

\clearpage

\end{document}